\documentclass{amsproc}

\usepackage[english]{babel}

\newtheorem{theorem}{Theorem}[section]

\theoremstyle{definition}

\theoremstyle{remark}

\numberwithin{equation}{section}

\usepackage[english]{babel}
\usepackage[T1]{fontenc}
\usepackage{hyperref}

\usepackage{amssymb}
\usepackage{eucal}

\begin{document}

\title[Space--adiabatic theory for
random--Landau Hamiltonian]{Space--adiabatic theory for
random--Landau Hamiltonian:
results and prospects}  

\author{Giuseppe De Nittis}

\address{LAGA - Universit\'{e} Paris 13 - Institut Galil\'{e}e, 93430 Villetaneuse, France.}

\email{denittis@math.univ-paris13.fr}

\thanks{The author would like to thank F. Klopp for many stimulating discussions.  Project supported by the grant ANR-08-BLAN-0261-01. }

\maketitle

The study of the  (integer) \emph{Quantum Hall Effect} (QHE) requires a careful analysis of the spectral properties of the $2D$, single-electron Hamiltonian
\begin{equation}\label{eq1}
H_{\Gamma,B}:=\left(-{\rm i}\partial_x-B\ y\right)^2+\left(-{\rm i}\partial_y+B\ x\right)^2+V_\Gamma(x,y)
\end{equation}
where $H_B:=H_{\Gamma,B}-V_\Gamma$ is the usual \emph{Landau Hamiltonian} (in symmetric gauge) with \emph{magnetic field} $B$ and $V_\Gamma$ is a  $\Gamma\equiv{\mathbb Z}^2$ \emph{periodic potential} which models the electronic interaction
with   a crystalline structure. Under usual conditions (e.g., $V_\Gamma\in L^2_{\rm loc}({\mathbb R}^2)$) the Hamiltonian \eqref{eq1} is self-adjoint on a suitable subdomain of $L^2({\mathbb R}^2)$.
A direct analysis of the fine spectral properties of \eqref{eq1} is extremely difficult and one needs resorting to  simpler effective models hoping to capture (some of) the main physical features in suitable physical regimes. 

\subsection*{Weak magnetic field limit}
The regime  $B\ll1$ is very interesting since it is easily accessible to experiments. The common lore, (cf.  works of  R.~Peierls, 
P.~G.~Harper and D.~Hofstadter), says that the   ``local description'' of the spectrum of \eqref{eq1} is ``well approximated''  by the spectrum of the  \emph{Hofstadter} (effective) \emph{model}
\begin{equation}\label{eq2}
\big(H^{(B)}_{\rm Hof}\xi\big)_{n,m}:=e_{B}^{m}\ \xi_{n+1,m}+\overline{e_{B}^{m}}\ \xi_{n-1,m}+\overline{e_{B}^{n}}\ \xi_{n,m+1}+e_{B}^{n}\ \xi_{n,m-1}
\end{equation}
with $\{\xi_{n,m}\}\in\ell^2({\mathbb Z}^2)$ and $e_{B}^{m}:={\rm e}^{{\rm i}2\pi mB}$.

The above discussion leads to the following questions:
\begin{enumerate}
\item[Q.1)] In what mathematical sense are ${\rm Spec}\big(H_{\Gamma,B}\big)$ and ${\rm Spec}\big(H^{(B)}_{\rm Hof}\big)$ ``locally equivalent''?
\item[Q.2)] What is the relation between the ``effective'' dynamics induced by $H^{(B)}_{\rm Hof}$ and the ``true'' dynamics induced by $H_{\Gamma,B}$?
\end{enumerate}
A third question concerns the r\^ole of the disorder in the explanation of the QHE. Indeed, the introduction of a \emph{random potential} $V_\omega$ (e.g., an \emph{Anderson potential}) in \eqref{eq1}, leading to
\begin{equation}\label{eq3}
H_{\Gamma,B,\omega}:=H_{\Gamma,B}+V_\omega,
\end{equation}
is essential in order to explain the emergence of the quantum Hall plateaus. Then:
\begin{enumerate}
\item[Q.3)] In presence of disorder is it still  possible to derive a  ``simplified''
 (i.e., effective) model for    
$H_{\Gamma,B,\omega}$ which encodes the (main) spectral and dynamical properties of the full model? 
\end{enumerate}

To answer to questions Q.1) and Q.2) one  needs to prove the so-called \emph{Peierls substitution}. This is an old problem which dates back to the works of J.~Bellissard \cite{bellissard-89}  and B.~Helffer and  J.~Sj{\"o}strand \cite{helfer-sjostrand-89}. However, these works  provide only a partial answer to Q.1) (\emph{local isospectrality}) and no answer for Q.2). A complete solution has been given only recently by the author and M. Lein in \cite{denittis-lein-11}.  In this paper a strong version of the Peierls substitution is derived
by means of a joint application of the \emph{Space-adiabatic perturbation theory} (SAPT) developed by G.~Panati, H.~Spohn and S.~Teufel \cite{panati-spohn-teufel-03} and the \emph{magnetic Weyl quantization}
developed by M.~M\u{a}ntoiu and R.~Purice \cite{mantoiu-purice-04}. The main result derived in \cite{denittis-lein-11} can be stated as follows:
\begin{theorem}\label{teo1}
Assume the existence of a $S\subset {\rm Spec}\big(H_{\Gamma,B=0}\big)$ separated from the rest of the spectrum ${\rm Spec}\big(H_{\Gamma,B=0}\big)\setminus S$ by  gaps\footnote{This assumption can be relaxed by introducing the notion of \emph{adiabatically decoupled} energy subspace, cf. \cite{panati-spohn-teufel-03} or \cite{denittis-lein-11}. }. Then:
\begin{enumerate}
\item[(i)] Associated to $S$ there exists an an orthogonal projection $\Pi_B$ in $L^2({\mathbb R}^2)$ such that for any $N\in{\mathbb N}$ 
\begin{equation}\label{eq5}
\big\|[H_{\Gamma,B};\Pi_B]\big\|\leqslant C_N\ B^N\qquad\text{if}\qquad B\to0
\end{equation}
where $C_N>0$ are suitable constants. The space ${\rm Ran}\ \Pi_B\subset L^2({\mathbb R}^2)$
 is called \emph{almost-invariant} subspace.
\item[(ii)] There exists a \emph{reference} Hilbert space $\CMcal{H}_{\rm r}$ ($B$-independent), an \emph{effective} (bounded) \emph{operator} $H_B^{\rm eff}$ on $\CMcal{H}_{\rm r}$ and a unitary operator $U_B:{\rm Ran}\ \Pi_B\to \CMcal{H}_{\rm r}$ such that for any $N\in{\mathbb N}$
\begin{equation}\label{eq6}
\big\|\big(
{\rm e}^{{\rm i} t H_{\Gamma,B}}-U^{-1}_B\ {\rm e}^{{\rm i} t H_B^{\rm eff}}\ U_B\big)\Pi_B\big\|\leqslant C_N\ B^N\ |t|\qquad\text{if}\qquad B\to0.
\end{equation}
\item[(iii)] If $S$ corresponds to a single Bloch energy band $E_\ast$ for the periodic operator $H_{\Gamma,B=0}$, then  $\CMcal{H}_{\rm r}\equiv\ell^2({\mathbb Z}^2)$. Moreover if the dispersion law for $E_\ast$ can be approximated  as
$E_\ast(k_1,k_2)=2\cos(k_1)+2\cos(k_2)+Bf(k_1,k_2)$, with $k_1$ and $k_2$ the Bloch momenta, then
\begin{equation}\label{eq7}
 H_B^{\rm eff}=H^{(B)}_{\rm Hof}+\CMcal{O}(B)\qquad\text{if}\qquad B\to0.
\end{equation}
\end{enumerate}
\end{theorem}
Theorem \ref{teo1} implies the  following answers for Q.1) and Q.2): $\Pi_BH_{\Gamma,B}\Pi_B$ and $H^{(B)}_{\rm Hof}$ are unitarily equivalent up to an error which goes to zero if $B\to0$ (\emph{asymptotic unitary equivalence}); the dynamics generated by $H^{(B)}_{\rm Hof}$ approximates the dynamics generated by $\Pi_BH_{\Gamma,B}\Pi_B$ up to a small error
over any macroscopic time-scale $t\in[0,T]$.

Question Q.3) suggests to combine SAPT-techniques with the randomness induced by $V_\omega$. However, one of the main ingredients of SAPT is the separation in fast and slow degrees of freedom induced by the periodic structure of $H_{\Gamma,B=0}$. This separation (mathematically highlighted by means of a Bloch-Floquet transform) identifies the fast part of the dynamics with the one inside the fundamental cell of $\Gamma$. The slow part is related to the motion at the boundary of adjacent cells and is controlled by the slow variation of the Bloch momenta induced by the weak, but non-zero, magnetic field $B\ll 1$. In order to include 
$V_\omega$ in this schema, one needs to assume that the randomness perturbs the periodic structure 
on a scale larger that the  typical length of the crystal and which becomes larger and larger when $B\to0$. In other words  SAPT-tecniques are compatible only with $B$-dependent random potentials of  type
\begin{equation}\label{eq8}
V_{\omega,B}(x,y):=w_\omega\big(B^{-1}x,B^{-1}x\big)
\end{equation}
with $w_\omega$ suitable random variables. If order to overcome the quite unphysical restriction \eqref{eq8}  one has to replace the usual Bloch-Floquet transform with some non-commutative extension. An hint in  this direction is provided by the  Bellissard's idea of replacing the Bloch-Floquet decomposition with the non-commutative notion of crossed product $C^\ast$-algebra \cite{Bellissard-Baldes-Elst-94}.

\subsection*{Strong magnetic field limit}
The opposite regime of a strong  magnetic field $B\gg1$ (accessible to experiments by means of optical lattices)  is mathematically easier to treat. In this regime, the dominant terms for the ``renormalized'' Hamiltonian $B^{-1}\ H_{\Gamma,B}$
 turns out to be a harmonic oscillator which fixes the energy threshold (Landau level). Under the assumption $V_\Gamma(x,y)\simeq 2\cos(x)+2\cos(y)$, the first relevant term for the asymptotic description of the spectral properties of  \eqref{eq1} is given by the \emph{Harper} (effective) \emph{model}
\begin{equation}\label{eq10}
\big(H^{(B)}_{\rm Har}\xi\big)(s):=\xi\big(s+B^{-1}\big)+\xi(s-B^{-1})+2\cos(2\pi s)\xi(s),\qquad \xi\in L^2({\mathbb R}).
\end{equation}
 The effective operator \eqref{eq10}  was firstly derived by M.~Wilkinson (1987). However, a rigorous (asymptotic) unitary derivation of 
 $H^{(B)}_{\rm Har}$ from $H_{\Gamma,B}$, in the spirit of the Theorem \ref{teo1}, has been established only recently
in \cite{denittis-panati-10}. Moreover, the proof can be successfully extended to the case of a random potential $V_\omega$ generalizing  a perturbative technique developped in \cite{eckstein-09}.



\begin{thebibliography}{10}

\bibitem{bellissard-89}
J.~V. Bellissard.
\newblock {C*-algebras in solid state physics: 2D electrons in uniform magnetic
  field}.
\newblock In: {\em Operator Algebras and Applications Vol. 2: Mathematical Physics and Subfactors}, E.~Evans et al. editors,  
{\em London Mathematical Society Lecture Note Series vol. 136}, Cambridge
University Press, 49--76 (1989).


\bibitem{Bellissard-Baldes-Elst-94}
J.~V. Bellissard, H.~{Schulz-Baldes},  A.~{van~Elst}.
\newblock {The noncommutative geometry of the quantum Hall effect}.
\newblock {\em J. Math. Phys.}
 {\bf 35}, 5373--5471 (1994).


\bibitem{denittis-lein-11}
G.~De~Nittis, M.~Lein.
\newblock {Applications of Magnetic $\Psi$-DO Techniques to Space-adiabatic Perturbation Theory}.
\newblock {\em Rev. Math. Phys.}
 {\bf 23} (3),  233--260 (2011).


\bibitem{denittis-panati-10}
G.~De~Nittis, G.~Panati.
\newblock {Effective models for conductance in magnetic fields: derivation of
  Harper and Hofstadter models}.
\newblock arXiv:1007.4786v1 [math-ph]. 


\bibitem{eckstein-09}
A.~Eckstein. 
\newblock {Unitary reduction for the two-dimensional Schr\"{o}dinger operator with strong magnetic field}.
\newblock {\em Mathematische Nachrichten}
 {\bf 282},   (2009).

\bibitem{helfer-sjostrand-89}
B.~Helffer, J.~Sj{\"o}strand.
\newblock {{\'E}quation de Schr{\"o}dinger avec champ magn{\'e}tique et
{\'e}quation de Harper}.
\newblock In {\em {Schr{\"o}dinger Operators (Proceedings of the Nordic Summer
School in Mathematics, Sandbjerg Slot, Denmark, 1988)}}, 
{\em Lecture Notes in Physics}, vol. 345, Springer,  118--197 (1989).

\bibitem{mantoiu-purice-04}
M.~M\u{a}ntoiu, R.~Purice.
\newblock {The Magnetic Weyl Calculus}
\newblock {\em J. Math. Phys.}
 {\bf 45}(4),  233--260 (2004).


\bibitem{panati-spohn-teufel-03}
G.~Panati, H.~Spohn, S.~Teufel.
\newblock {Effective dynamics for Bloch electrons: Peierls substitution and beyond}.
\newblock {\em  Comm. Math. Phys.} {\bf 242}, 547--578 (2003).











\end{thebibliography}
\end{document}